\documentclass[twoside,twocolumn,9pt]{article}

\usepackage{extsizes}
\usepackage[super,sort&compress,comma]{natbib} 
\usepackage[version=3]{mhchem}
\usepackage[left=1.5cm, right=1.5cm, top=1.785cm, bottom=2.0cm]{geometry}
\usepackage{balance}
\usepackage{times,mathptmx}
\usepackage{sectsty}
\usepackage{lastpage}
\usepackage[format=plain,justification=justified,singlelinecheck=false,font={stretch=1.125,small,sf},labelfont=bf,labelsep=space]{caption}
\usepackage{float}
\usepackage{fancyhdr}
\usepackage{fnpos}
\usepackage[english]{babel}
\addto{\captionsenglish}{%
  
}
\usepackage{array}
\usepackage{droidsans}
\usepackage{charter}
\usepackage[T1]{fontenc}
\usepackage[utf8]{inputenc}
\usepackage[usenames,dvipsnames]{xcolor}
\usepackage{setspace}
\usepackage[compact]{titlesec}
\usepackage{hyperref}

\setcitestyle{super}

\newcommand*{\citen}[1]{%
  \begingroup
    \romannumeral-`\x 
    \setcitestyle{numbers}%
    \cite{#1}%
  \endgroup   
}

\usepackage{graphicx}
\graphicspath{{graphs//}}
\usepackage{subfigure}
\definecolor{cream}{RGB}{222,217,201}

\begin{document}

\pagestyle{fancy}
\thispagestyle{plain}
\fancypagestyle{plain}{

\renewcommand{\headrulewidth}{0pt}
}

\makeFNbottom
\makeatletter
\renewcommand\LARGE{\@setfontsize\LARGE{15pt}{17}}
\renewcommand\Large{\@setfontsize\Large{12pt}{14}}
\renewcommand\large{\@setfontsize\large{10pt}{12}}
\renewcommand\footnotesize{\@setfontsize\footnotesize{7pt}{10}}
\makeatother

\renewcommand{\thefootnote}{\fnsymbol{footnote}}
\renewcommand\footnoterule{\vspace*{1pt}%
\color{cream}\hrule width 3.5in height 0.4pt \color{black}\vspace*{5pt}} 
\setcounter{secnumdepth}{5}

\makeatletter 
\renewcommand\@biblabel[1]{#1}            
\renewcommand\@makefntext[1]%
{\noindent\makebox[0pt][r]{\@thefnmark\,}#1}
\makeatother 
\renewcommand{\figurename}{\small{Fig.}~}
\sectionfont{\sffamily\Large}
\subsectionfont{\normalsize}
\subsubsectionfont{\bf}
\setstretch{1.125} 
\setlength{\skip\footins}{0.8cm}
\setlength{\footnotesep}{0.25cm}
\setlength{\jot}{10pt}
\titlespacing*{\section}{0pt}{4pt}{4pt}
\titlespacing*{\subsection}{0pt}{15pt}{1pt}

\fancyfoot{}
\fancyfoot[RO]{\footnotesize{\sffamily{1--\pageref{LastPage} ~\textbar  \hspace{2pt}\thepage}}}
\fancyfoot[LE]{\footnotesize{\sffamily{\thepage~\textbar\hspace{3.45cm} 1--\pageref{LastPage}}}}
\fancyhead{}
\renewcommand{\headrulewidth}{0pt} 
\renewcommand{\footrulewidth}{0pt}
\setlength{\arrayrulewidth}{1pt}
\setlength{\columnsep}{6.5mm}
\setlength\bibsep{1pt}

\makeatletter 
\newlength{\figrulesep} 
\setlength{\figrulesep}{0.5\textfloatsep} 

\newcommand{\topfigrule}{\vspace*{-1pt}%
\noindent{\color{cream}\rule[-\figrulesep]{\columnwidth}{1.5pt}} }

\newcommand{\botfigrule}{\vspace*{-2pt}%
\noindent{\color{cream}\rule[\figrulesep]{\columnwidth}{1.5pt}} }

\newcommand{\dblfigrule}{\vspace*{-1pt}%
\noindent{\color{cream}\rule[-\figrulesep]{\textwidth}{1.5pt}} }

\makeatother

\twocolumn[
  \begin{@twocolumnfalse}
\vspace{3cm}
\sffamily
\begin{tabular}{m{4.5cm} p{13.5cm} }

& \noindent\LARGE{\textbf{High-resolution resonance-enhanced multiphoton photoelectron circular dichroism}} \\
\vspace{0.3cm} & \vspace{0.3cm} \\

 & \noindent\large{Alexander Kastner,$^{a}$ Greta Koumarianou,$^{b}$ Pavle Glodic,$^{b}$ Peter C. Samartzis,$^{b}$ Nicolas Ladda,$^{a}$ Simon T. Ranecky,$^{a}$ Tom Ring,$^{a}$  Vasudevan Sudheendran,$^{a}$ Constantin Witte,$^{a}$ Hendrike Braun,$^{a}$ Han-Gyeol  Lee,$^{a}$  Arne Senftleben,$^{a}$ Robert Berger,$^{c}$ G. Barratt Park,$^{d,e}$ Tim Schäfer,$^{d,e}$ $^\dag$ and Thomas Baumert$^{a}$ } \\

& \noindent\normalsize{
Photoelectron circular dichroism (PECD) is a highly sensitive enantiospecific spectroscopy for studying chiral molecules in the gas phase using either single-photon ionization or multiphoton ionization. In the short pulse limit investigated with femtosecond lasers, resonance-enhanced multiphoton ionization (REMPI) is rather instantaneous and typically occurs simultaneously via more than one vibrational or electronic intermediate state due to limited frequency resolution. In contrast, vibrational resolution in the REMPI spectrum can be achieved using  nanosecond lasers. In this work, we follow the high-resolution approach using a tunable narrow-band nanosecond laser to measure REMPI-PECD through distinct vibrational levels in the intermediate 3s and 3p Rydberg states of fenchone. We observe the PECD to be essentially independent of the vibrational level. This behaviour of the chiral sensitivity may pave the way for enantiomer specific molecular identification in multi-component mixtures: one can specifically excite a sharp, vibrationally resolved transition of a distinct molecule to distinguish different chiral species in mixtures.

}
\end{tabular}

 \end{@twocolumnfalse} \vspace{0.6cm}

  ]

\renewcommand*\rmdefault{bch}\normalfont\upshape
\rmfamily
\section*{}
\vspace{-1cm}


\footnotetext{\textit{$^{a}$~Universität Kassel, Heinrich-Plett-Str. 40, 34132 Kassel, Germany. }}
\footnotetext{\textit{$^{b}$~Institute of Electronic Structure and Lasers, Foundation for Research and Technology - Hellas (FORTH), P.O. Box 1527, 71110 Heraklion, Greece. }}
\footnotetext{\textit{$^{c}$~Fachbereich Chemie, Philipps-Universität Marburg, Hans-Meerwein-Straße 4, 35032 Marburg, Germany.}}
\footnotetext{\textit{$^{d}$~Georg-August-Universität Göttingen, Tammannstr. 6, 37077 Göttingen, Germany. }}
\footnotetext{\textit{$^{e}$~Max Planck Institut für biophysikalische Chemie, Am Fassberg 11, 37077 Göttingen, Germany.}}


\footnotetext{\dag email \href{mailto:tschaef4@gwdg.de}{tschaef4@gwdg.de} }


\section{Introduction}

%
%

Enantiomer-specific spectroscopy of chiral molecules in the gas phase \cite{Zehnacker.2010} is a dynamically evolving research field, where techniques like e.g. microwave three-wave mixing,\cite{Patterson.2013, Patterson.2013a, Eibenberger.2017} chiral-sensitive high harmonic generation,\cite{Cireasa.2015, Woerner.2018, Neufeld.2019} controlled enantioselective orientation with an optical centrifuge,\cite{Milner.2019} or Coulomb explosion imaging for absolute configuration determination\cite{Pitzer.2013, Herwig.2013, Pitzer.2016, Pitzer.2016b, Pitzer.2018} have been developed in recent years. Chiral signatures are also present in both the photoion and photoelectron signal when chiral molecules are ionized with circularly polarized light. Circular dichroism (CD) in the ion yield has been investigated using resonance enhanced multi-photon ionization (REMPI) in combination with mass spectrometry.\cite{Boesl.2006, Compton.2006, Breuning.2009, Lepelmeier.2016, Boesl.2013, Horsch.2012} Coulomb explosion imaging techniques have recently allowed transient chirality to be investigated in the molecular frame.\cite{Fehre.2018} In contrast to REMPI-based CD of the photoion yield, which arises from weak interference effects between electric and magnetic dipole transition moments, the chiral signature of the photoelectron angular distribution after laser ionization originates from an electric dipole interaction alone. In 1976, it was theoretically predicted that the photoionization of optically active molecules by circularly polarized light would give rise to a substantial forward-backward asymmetry in the angular distribution of the photoelectrons.\cite{Ritchie.1976} This effect, known as photoelectron circular dichroism (PECD), was first demonstrated by single-photon ionization of bromocamphor using a synchrotron source in 2001.\cite{Bowering.2001} Subsequently, PECD was also demonstrated on other bicyclic ketones using pulsed femtosecond table-top laser systems to drive 2+1 REMPI.\cite{Lux.2012, Lehmann.2013} The angular distribution of photoelectrons was measured using velocity map imaging.\cite{Chandler.1987, Eppink.1997, Chandler.2017} Multiphoton PECD has also been observed for methyloxirane\cite{Fanood.2014} and limonene.\cite{Fanood.2015b, Beaulieu.2016a} Because PECD has been observed when using single-photon ionization out of an achiral core-shell initial orbital,\cite{Hergenhahn.2004, Ulrich.2008} it has been interpreted as being largely a final state effect, strongly influenced by the long-range scattering potential.\cite{Garcia.2013} In contrast, PECD has also been observed for photoelectrons with high kinetic energy above 500 eV resulting from a Fano interference.\cite{Hartmann.2019}
PECD is sensitive to the final vibrational level of the cation,\cite{Garcia.2013, Garcia.2017, Fanood.2018}  conformation of the molecule,\cite{Fanood.2018, Turchini.2017} and to subcycle interactions with tailored bichromatic light fields.\cite{Demekhin.2018, Rozen.2019}  \\ \indent 
The underlying electric dipole interaction makes PECD very sensitive to small changes in the amount of enantiomeric excess (e.e.) and consequently the technique shows promise as a highly sensitive analytical tool\cite{Lux.2012, Lux.2015, Lehmann.2013, Janssen.2017, Nahon.2016, Comby.2018} exhibiting sub one percent sensitivity. \cite{Kastner.2016} It is possible to measure enantiomeric excesses in multi-component mixtures by recording mass-selected PECD spectra with electron image-ion mass coincidence methods.\cite{Fanood.2015} This approach requires significant experimental effort so that alternative methods are desirable. For instance, PECD in high-resolution REMPI in molecular beams using nanosecond laser systems could be a promising spectroscopic approach as speculated previously by Janssen \textit{et al.} \cite{Janssen.2014} The technique permits vibrational resolution, so that different chiral molecules in mixtures can be distinguished by their vibrational features in the REMPI spectra. This has already been demonstrated in REMPI-CD experiments that have taken advantage of high laser resolution to observe sign changes when investigating REMPI-CD via different vibrational levels of the intermediate electronic state\cite{Lepelmeier.2017} or to distinguish different conformers.\cite{Hong.2014} PECD spectra belonging to a certain molecular species can also be recorded by ionizing via narrow, unambiguously assigned REMPI transitions. Of course, for these kind of experiments rotational and vibrational degrees of freedom should be cooled, which can be efficiently achieved using supersonic molecular beams.  \\ \indent
Recently, a femtosecond pump-probe setup was used to observe picosecond dynamics in the 3s Rydberg intermediate state of fenchone, one of the typical benchmark molecules for PECD measurements.\cite{Comby.2016} The PECD value was found to vary on a picosecond timescale. However, due to the large frequency bandwidth of the femtosecond laser pulse (30 meV), no vibrational resolution could be achieved. In other studies on fenchone, the dependence on intermediate electronic state and photoelectron kinetic energy has been investigated.\cite{Beaulieu.2016a, Kastner.2017} This information provides valuable benchmarks for emerging theoretical descriptions of multiphoton PECD,\cite{Goetz.2017, Lein.2014, Demekhin.2015, Lehmann.2013, Mueller.2018, Demekhin.2020} and coherent control approaches.\cite{Koch.2018}\\ \indent
In this contribution we demonstrate partially vibrationally resolved multiphoton PECD of fenchone, obtained by high-resolution 2+1 REMPI via intermediate Rydberg states with a nanosecond pulsed dye laser system. Earlier we observed nanosecond-pulse PECD without vibrational resolution with the help of a frequency tripled Nd:YAG laser.\cite{Kastner.2019} Experiments are supported by theoretical modeling of the electronically excited states' vibrational level structure. In the low-energy region of the 3s state, the features observed in a previous 2+1 REMPI spectrum\cite{Kastner.2017} can be assigned to single vibrational eigenstates. As the photon energy is increased, the density of vibrational states increases rapidly and intramolecular vibrational redistribution (IVR) dynamics of the vibrational wavepacket that is excited on the 3s Rydberg state become important. We do not observe significantly different PECD values when ionizing fenchone either via a resolved eigenstate or via a wavepacket made up of many strongly interacting vibrational levels. Hence, the vibrational character of the intermediate 3s Rydberg state does not substantially influence the PECD of fenchone after 2+1 REMPI.  
This work is the first demonstration of PECD employing tunable dye laser systems providing vibrational resolution. The approach presented here based on high-resolution REMPI spectroscopy has the potential to facilitate enantiomer-specific analytical measurements of multi-component mixtures.

\section{Methods}
\subsection{Experimental Setup and Data Evaluation}
\label{sec:ExpSetup}

\begin{figure}[tb]
\centering
\includegraphics[width=0.85\linewidth]{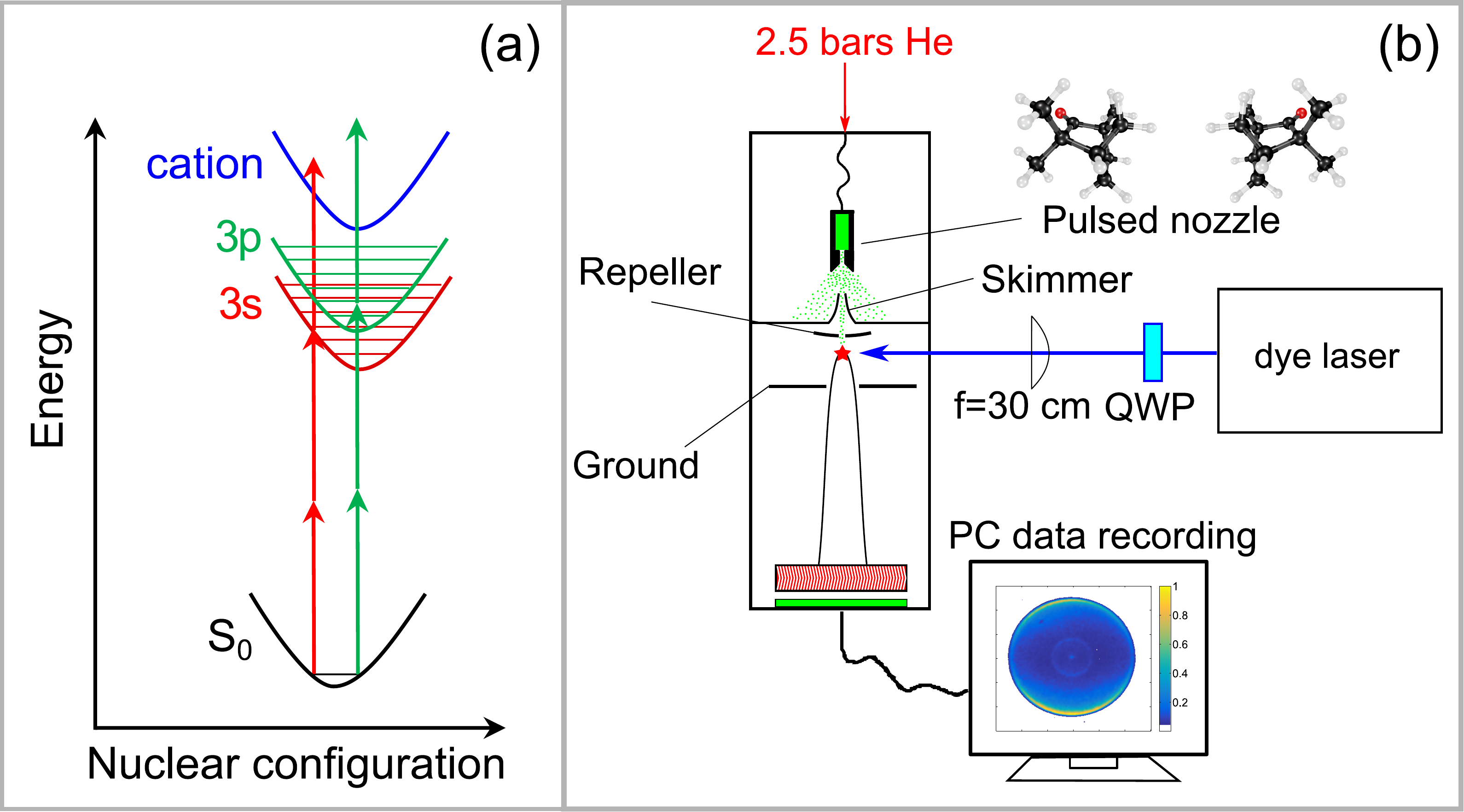}
\caption{\label{fig:Expsetup}
(a) 2+1 Resonance-enhanced multiphoton ionization (REMPI) scheme: Two photons are necessary to excite vibrational levels of the 3s or 3p Rydberg states while another photon ionizes the molecule. (b) Experimental setup for the vibrationally resolved PECD experiment using a high resolution dye laser. A detailed description is given in the text.  
}
\end{figure}

The experimental setup is shown in Figure \ref{fig:Expsetup}. A commercial dye laser (Lambda Physik LPD 3000) was pumped by a Lambda Physik LPX 315i XeCl excimer laser (10 Hz, 308 nm, 25 ns). The output of the dye laser was tuned over the wavelength region 375--420 nm using the laser dyes QUI, PBBO, and Exalite 416. The pulse energy was typically 3--6 mJ and the bandwidth was 0.1 cm$^{-1}$. The approximate laser wavelength was measured using an Avantes AvaSpec 3648 spectrometer (324\textendash 482 nm range, about 0.04 nm resolution). A more precise calibration was obtained by comparing the peak positions in the REMPI spectrum to those that were measured previously,\cite{Kastner.2017} which were calibrated to an accuracy of $\pm0.03$ cm$^{-1}$ using a high-precision wavemeter (HighFinesse, WS7). 

The high-resolution spectrum of fenchone\cite{Kastner.2017} recorded with linearly polarized light shows the partially resolved vibrational structure of 3s and 3p electronic states (shown in blue in Figure \ref{fig:REMPIvib}). It agrees in the lower wavenumber range nicely with the spectrum reported by Driscoll \textit{et al.},\cite{Driscoll.1991} who recorded the 3s region. Different electronic intermediate states can be populated by the two-photon transition. The two lowest-lying electric dipole-allowed Rydberg transitions lead to electronic intermediates having 3s and 3p Rydberg character.\cite{Pulm.1997} The transition from these intermediates to the continuum was observed to be  governed by a $\Delta v=0$ propensity rule in a previous femtosecond experiment.\cite{Kastner.2017} \\ \indent
The laser beam polarization was initially linear and oriented in the plane of the detector. An achromatic quarter-wave plate (B.Halle, 300\textendash 470 nm achromatic with air gap) was used to convert linearly polarized (LIN) to left circularly polarized (LCP) or right circularly polarized (RCP) light. The wavelength scan was split into three parts corresponding to the dyes used. The quality of polarization was determined using a Glan-Laser polarizer (ThorLabs GL10) and a power meter (Ophir Nova II). The circularity of polarization is expressed via the Stokes |S$_3$| parameter and has been measured for each dye at different wavelength settings. Except for one measurement point at 396 nm, where the laser power was unstable, |S$_3$| was well above 96\%. No correction of the experimental data with respect to the |S$_3$| value was performed. \\ \indent
The enantiopure (S)-($+$)- and (R)-($-$)-fenchone samples were purchased from Sigma-Aldrich with a specified purity of 99.2\% (fenchone) and used without further purification. The enantiomeric excess (e.e.) as measured by gas chromatography was 99.9\% for (S)-($+$)-fenchone and 84\% for \mbox{(R)-($-$)-fenchone}. \cite{Kastner.2016} 
Liquid fenchone, soaked into an inert cotton wad, was evaporated in a heatable sample compartment near the tip of a home-built pulsed solenoid valve nozzle, which is based on the Even-Lavie design\cite{Even.2000} and has been described in detail previously.\cite{Formaldehyde61Paper}
The fenchone sample was heated to about 100$^\circ$C, and 2.5 bars of He was used as a backing gas. 
The supersonic beam was singly skimmed before entering the detection chamber, where it is intersected by the laser beam at a 90$^\circ$ angle. \\ \indent
An \mbox{$f = 300 $ mm} plano-convex quartz lens was used to focus the laser beam into the interaction region of an imaging photoelectron spectrometer. The focal distance was chosen to avoid power broadening of the absorption lines. The photoelectron spectrometer, which has been explained in detail in previous work,\cite{Papadakis.2006} uses velocity map-imaging (VMI) as demonstrated by Eppink and Parker\cite{Eppink.1997, Chandler.2017} with the ion focusing lens included in the geometry of the repeller and ground plates. See Fig. \ref{fig:Expsetup} b).
The VMI provides an energy resolution of about $\Delta E/E \approx 5\%$. 
Energy calibration of the VMI was performed using ionization of Xe to the $^2P_{1/2}$ and $^2P_{3/2}$ continuum states via a four-photon transition driven by the third harmonic of a Nd:YAG laser. 
The VMI can be used to project either the photoelectrons or the photoions onto a home-built imaging detector composed of multi-channel plates (BASPIK) with a 50 mm diameter and a P47 phosphor screen (Proxivision). Three dimensional photoelectron angular distributions (PADs) are projected onto the detector and measured as PAD images using a 1.4 million pixel CCD camera (Unibrain Fire-i 702b). The VMI voltage was set such that photoelectrons with kinetic energy up to about 4.1 eV could be measured. Alternatively, ion TOF mass spectra can be recorded on an oscilloscope (Hameg model 1507-3; 150 MHz, 200 MS/s) via a capacitively coupled output. 
At each wavelength, photoelectron images for LIN, LCP, and RCP laser light polarization were averaged for 1800 laser pulses ($\sim$ 3 minutes). 
The PECD image was calculated by subtracting the averaged RCP PAD image from the averaged LCP PAD image. \\ \indent
The original three-dimensional photoelectron distribution is reconstructed using an Abel inversion routine, where two different algorithms have been used. The LIN PADs were evaluated by a polar onion peeling\cite{Roberts.2009} algorithm written in MATLAB. The polar onion peeling algorithm runs on the pixel grid provided by the camera. The LCP and RCP PADs were evaluated using a C$++$ based pBasex algorithm,\cite{Garcia.2004} which was kindly provided by G. A. Garcia from SOLEIL. Details on the evaluation can be found in the Supporting Information. We used a Legendre polynomial basis function set that includes both odd and even order polynomials up to the 6$^\text{th}$ order following Yang's theorem.\cite{Yang.1948}    
Using the pBasex algorithm, different peaks in the photoelectron spectrum can be evaluated separately. The chiral signature is contained in the odd-order Legendre polynomial coefficients. The magnitude of PECD can be derived by computing a sum over the odd-order coefficients $c_i$ normalized to the total signal $c_0$, denoted as linear PECD: \cite{Lux.2015}
\begin{equation}\label{LPECDpB}
\mathrm{LPECD} = \frac{1}{c_0} \left(2 c_1 - \frac{1}{2} c_3 + \frac{1}{4} c_5 \right).
\end{equation}
The LPECD value for a measurement is derived using a weighted average over the range of radii contained within the full width at half maximum (FWHM) of the peak signal. 

\subsection{Computational Methods}
Equilibrium structures and harmonic vibrational frequencies of fenchone in the electronic ground state and various Rydberg states were calculated on the approximate second order coupled cluster level (CC2) using the so-called resolution of the identity (RI) as implemented in the quantum chemistry program Turbomole.\cite{turbomole7.3} As in some of the calculations reported in previous work,\cite{Goetz.2017} we have chosen the cc-pVTZ basis set on all atoms except oxygen. For the latter we chose the t-aug-cc-pVTZ basis set. For C and H we chose the corresponding recommended RI basis sets (cbas, jkbas), for O we used the basis set recommended for d-aug-cc-pVTZ for fitting of the Coulomb and exchange terms (jkbas) and the original t-aug-cc-pVTZ basis also as RI basis set within the electron correlation treatment (cbas). The latter choice is not ideal, but resulted, in contrast to test calculations with the RI basis recommended for d-aug-cc-pVTZ, only in vertical excitation wavenumbers into the relevant Rydberg states that are systematically lower by about 46 cm$^{-1}$ to 50 cm$^{-1}$. Resulting structural changes in the electronic ground state were found to be negligible. Electrons in the 11 energetically lowest molecular orbitals, which are composed mostly of atomic 1s orbitals of oxygen and carbon, were kept frozen in the electron correlation treatment (frozen core approximation). Convergence criteria for the self-consistent field (SCF) energies were chosen to be $10^{-9}~E_\mathrm{h}$. Convergence criteria for the root mean square of the density matrix used in energy minimization and vibrational frequency calculations was on the order of $10^{-8}$. Ground state coupled cluster equations were iterated until the energy change between two cycles remained below $10^{-8}~E_\mathrm{h}$. The convergence threshold for the norm of residual vectors in subsequent linear response calculations of excitation energies was chosen on the $10^{-6}$ level. Convergence criteria for equilibrium structures were imposed such that the norm of analytic gradients of the energy with respect to displacements of the nuclei remained below $10^{-5}~E_\mathrm{h}~a_0^{-1}$ and the estimated energy change between two optimization steps was below $10^{-8}~E_\mathrm{h}$.  Franck--Condon profiles for excitations from the electronic ground state into the various Rydberg states were calculated with a development version of the hotFCHT program \cite{berger.1997a,jankowiak.2007,huh.2012proc} using both time-dependent and time-independent methodologies in the harmonic approximation. In the time-dependent approach, a Gaussian lineshape function with a full width at half maximum of $10~\mathrm{cm}^{-1}$ was chosen to mimic the approximate width of the 0$_0^0$ transition region of the 3s state. The time-independent approach was utilized with pre-screening criteria such that at least 97\% of the total integrated FC profile could be obtained. A maximum of five simultaneously excited normal modes was included for this purpose. The graining in the time-independent calculations was chosen to be $1~\mathrm{cm}^{-1}$. The calculated spectra are compared with
experiment in Figure 2.

\section{Results and Discussion}
\label{sec:Results}

\subsection{Resonance Enhanced Multiphoton Ionization}  

\begin{figure}[tb]
\centering
\includegraphics[width=\linewidth]{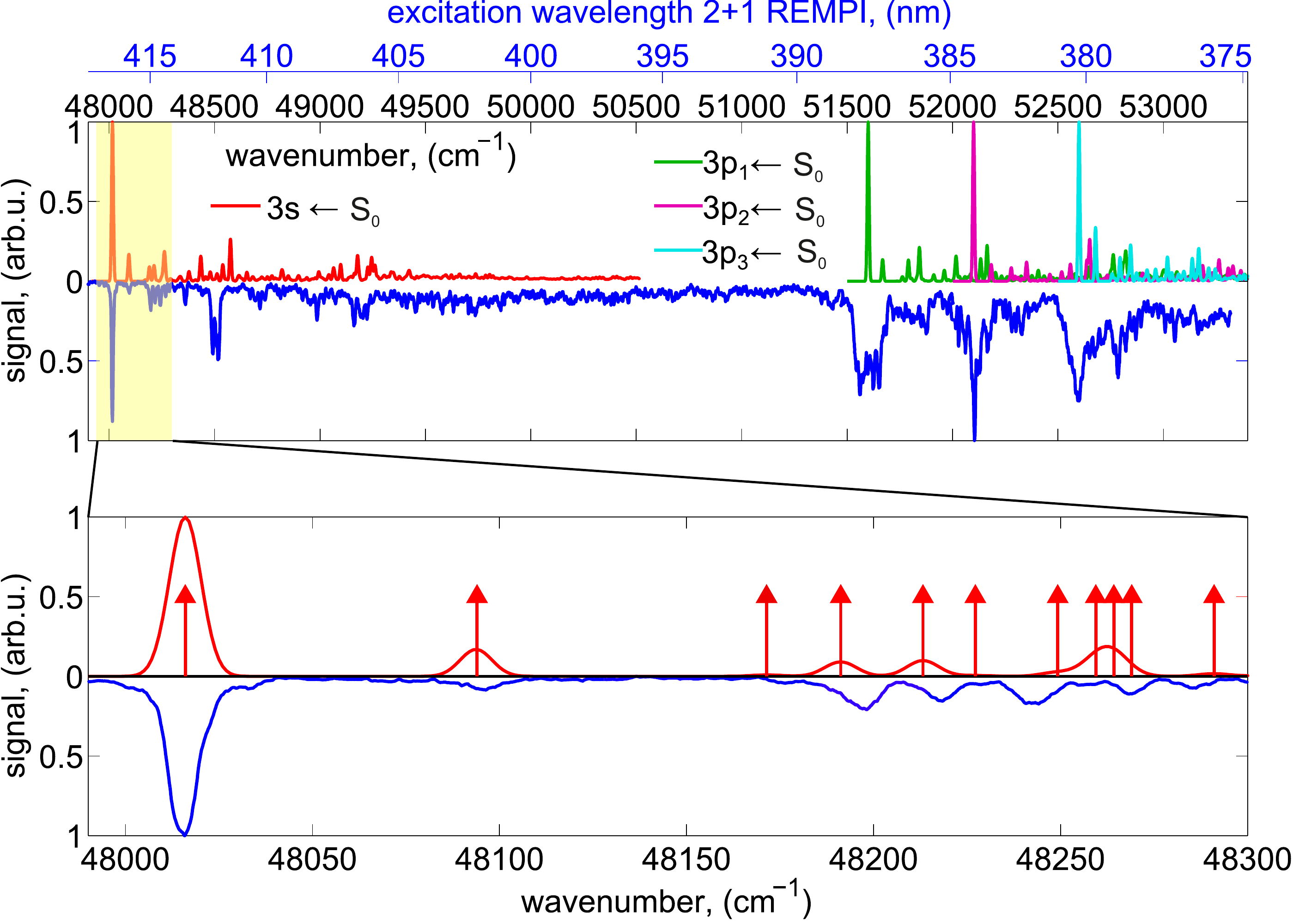} 
\caption{\label{fig:REMPIvib}
Experimental 2+1 REMPI spectrum recorded with linearly polarized light of fenchone involving the 3s $\leftarrow$ n and 3p $\leftarrow$ n transitions shown in blue. For comparison, the calculated Franck-Condon profiles for the transition to the 3s state (red) and the transitions to the three components of the 3p state, denoted here by 3p$_1$, 3p$_2$, and 3p$_3$ (green, purple, turquoise) are shown. The origin of the individual profiles have been shifted by about 4500 cm$^{-1}$ (namely 4348 cm$^{-1}$, 4257 cm$^{-1}$, 4466 cm$^{-1}$, 4746 cm$^{-1}$ for 3s, 3p$_1$, 3p$_2$, 3p$_3$, respectively)  to higher wavenumbers compared to the computed 0-0 transition wavenumbers. See text for more information. Intensities of the different FC profiles were scaled  such that all 0-0 transitions are normalized to the same intensity. In the bottom panel, an enlarged view of the region around the band origin of the 3s state, where the vibrational levels are well resolved, is shown. Transitions to individual vibrational eigenstates are indicated by blue arrows with their length being unrelated to the Franck-Condon factor.
}
\end{figure}

The 2+1 REMPI spectrum recorded with linearly polarized light shown in blue in Figure \ref{fig:REMPIvib} probes the excitation via the 3s state at energies $>$48000\,cm$^{-1}$ and the excitation via the 3p states at energies $>$51500\,cm$^{-1}$. Vibronic transitions are on average well separated below excitation energies of 48300\,cm$^{-1}$. In this region, the nanosecond laser (0.1 cm$^{-1}$ bandwidth) populates mostly single vibrational eigenstates during the REMPI process. 

With increasing wavenumber, however, the vibrational density of states grows rapidly, so that zero-order bright states are excited that can couple to a manifold of dark vibrational states located within the 0.1 cm$^{-1}$ laser bandwith leading at higher wavenumbers to IVR probably on a subpicosecond timesscale.\cite{Nesbitt.1996} Spectral lines at energies above 51500\,cm$^{-1}$ are significantly broadened indicating a fast dissipation channel out of the 3p Rydberg states. In the plots, the origins of the computed Franck--Condon profiles were shifted by about 4500 cm$^{-1}$ (namely 4348 cm$^{-1}$, 4257 cm$^{-1}$, 4466 cm$^{-1}$, 4746 cm$^{-1}$ for 3s, 3p$_1$, 3p$_2$, 3p$_3$, respectively)  to higher wavenumbers compared to the computed 0$_0^0$ transitions to match the experimentally observed spectrum. This corresponds only to one of several tentative assignments that appear possible for the three components of the 3p Rydberg state. The lowest fundamental vibration at 48090 cm$^{-1}$ is a delocalized butterfly motion of the ring systems. Vibrations between 48170 cm$^{-1}$ and 48300 cm$^{-1}$ are methyl torsional and skeletal modes. More details about the vibrations can be found in the Supporting Information.

\begin{figure}[tb]
\centering
\subfigure[\label{fig:PES}]{\includegraphics[width=0.49\linewidth]{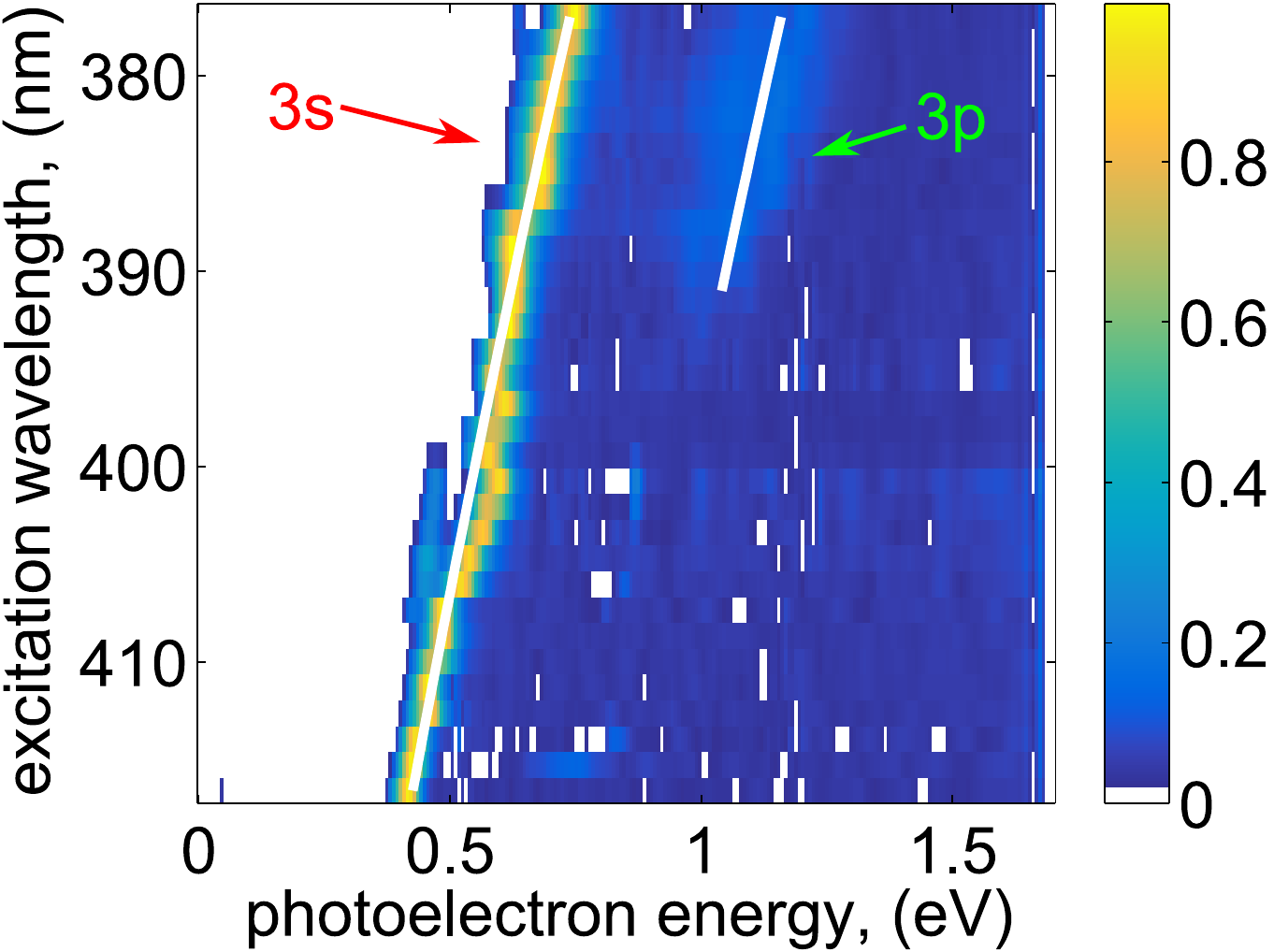}} \hfill
\subfigure[\label{fig:MassSpec}]{\includegraphics[width=0.49\linewidth]{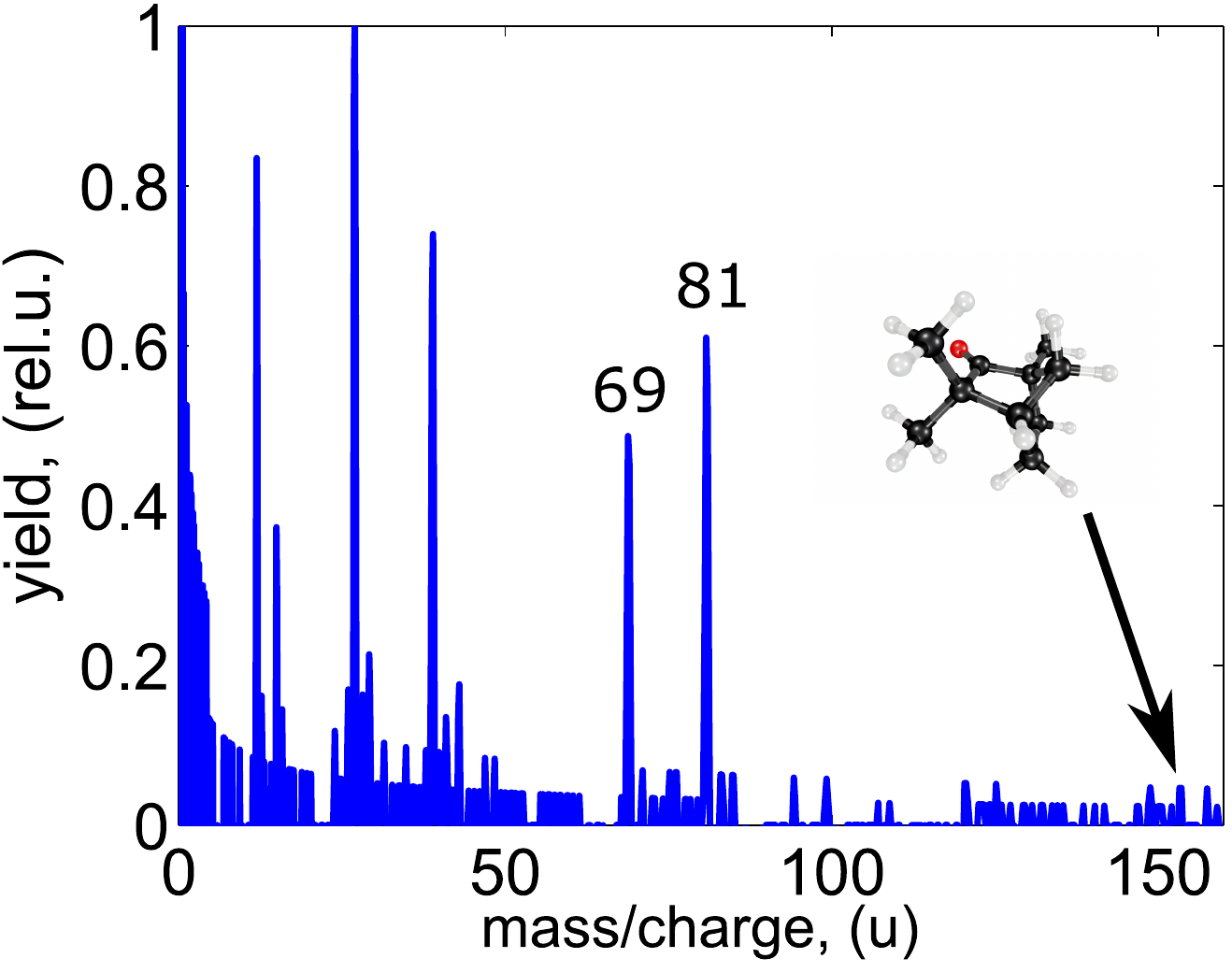}}
\caption{
(a) Wavelength-dependent PES derived from the LIN PADs for (S)-($+$)-fenchone after Abel-inversion using polar onion peeling.\cite{Roberts.2009} Every row represents a single measurement, where up to two distinct electronic intermediates were observed. The expected energy scaling with $\hbar\omega$ is indicated by the white lines. (b) Mass spectrum for (S)-($+$)-fenchone obtained at the 3s band origin at 416.57 nm. A detailed discussion can be found in the text.    
}
\end{figure}

We recorded photoelectron images at 27 ((R)-($-$)-fenchone) and 34 ((S)-($+$)-fenchone) selected excitation wavelengths between 375 nm and 420 nm for LIN, LCP, and RCP laser light. At wavelengths longer than 412 nm, the wavelength was tuned to the center of each resolved spectral feature in the REMPI spectrum. At the employed laser pulse energy no signal was observed off resonance. At wavelengths shorter than 412 nm, where the onset of a continuous unresolved background is observed, PAD images were recorded at 1 nm intervals. The photoelectron spectrum (PES) obtained for different excitation wavelengths of linearly polarized light are shown in Figure \ref{fig:PES}. Each PES contains a sharp, intense peak corresponding to ionization from the 3s intermediate state. At wavelengths shorter than about 389 nm a weak second contribution originating from excitation of the 3p intermediate states is observed in the PES. As the wavelength is scanned, the photoelectron peak position shifts by an amount equal to the change in the one-photon energy in agreement with findings in a previous femtosecond experiment.\cite{Kastner.2017} The scaling of photoelectron energy with excitation wavelength agrees well with $hc/\lambda-(IP-E_{\mathrm{3s,3p}})$ (shown as white lines in Figure \ref{fig:PES}). $IP=8.5$ eV is the adiabatic ionization energy of fenchone,\cite{Frost.1980, Kastner.2017} and $E_{\mathrm{3s}}=5.95$ eV and $E_{\mathrm{3p}}\approx 6.37$ eV are the electronic energies of the 3s and 3p intermediate states.\cite{Kastner.2017} The photoelectron energy scaling indicates essentially vibrational-energy conserving one-photon ionization out of the intermediate states which points to their potential energy surfaces being nearly parallel to the cationic one.

The intensity of the 3p peak relative to the 3s peak is significantly weaker than it is in the PES recorded in the same spectral range with femtosecond laser pulse ionization, where the contribution of the 3p state is larger than the contribution of the 3s state (see Figure 5 of reference \citen{Kastner.2017}). However, we observe a sharp increase in the total ion signal of about one order of magnitude when the 3p excitation threshold is reached. See excitation energies higher than 51500 cm$^{-1}$ in Figure \ref{fig:REMPIvib} and compare to the signal below 51500 cm$^{-1}$, which corresponds to transitions via the 3s state and is non-zero. This observation is consistent with internal conversion (IC) from the 3p state to the 3s state, leading to a shortened lifetime of the 3p state and hence to less ionization out of this state. Instead, molecules are ionized from the 3s state and emit photoelectrons carrying the energy signature of the 3s state (i.e. without the energy that has been converted to vibrational excitation in the 3s state). In general IC to the electronic ground state $S_0$ or to the energetically low-lying $S_1$ (n$\rightarrow$ $\pi^*$) excited state is also energetically possible. These states would be highly vibrationally excited after IC and are not observed in our experiment probably due to negligible Franck-Condon factors for ionization at the energy that is probed. However, keeping in mind that the total ion signal grows when exciting via the 3p state, and that the photoelectrons carry the energy signature from the 3s state, our observations suggest that 3p$\rightarrow$3s IC may be much more rapid than 3p$\rightarrow$S$_0$/S$_1$ IC due to the unfavorable Franck-Condon overlap for the latter processes. In contrast, when 2+1 REMPI via the 3p state is performed using a femtosecond laser pulse, the timescale of the experiment is too short for 3p$\rightarrow$3s IC to occur, and direct ionization out of the 3p state is observed more frequently.\cite{Kastner.2019} The intensity of the 3p Rydberg state peak in the photoelectron spectrum is therefore significantly higher than observed in the nanosecond laser experiment. \\ \indent
Pump-probe experiments conducted at $49800\pm200$~cm$^{-1}$ have shown that the 3s state undergoes internal conversion to the electronic ground state on a timescale of approximately 3.3 ps.\cite{Comby.2016} The width of the vibrationally resolved lines in the 48000--48500~cm$^{-1}$ region of our spectrum provides only a lower limit to the lifetime since the rotational structure is not resolved. Assuming a 10~K rotational temperature, and using the rotational constants reported in reference \citen{Loru.2016}, we obtain a simulated rotational contour with a FWHM of approximately 5~cm$^{-1}$, which depends by only a small amount on which component of the rank-two transition moment tensor is chosen. This value is only slightly narrower than the observed 9.1~cm$^{-1}$ FWHM of the $0_0^0$\,transition. The observed width corresponds to a lower limit of about 0.6~ps for the lifetime. This value is in reasonable agreement with the above mentioned lifetime of about 3.3~ps attributed to IC to the electronic ground state.\cite{Comby.2016} \\ \indent
The photoionization mass spectrum obtained at the band origin of the 3s state at 416.57 nm is shown in Figure \ref{fig:MassSpec}.
No parent ion is observed and the mass spectrum shows strong fragmentation of the molecules. Prominent fragments with largest mass-to-charge ratio have masses 69 and 81 amu. The observations made here are in agreement with previous nanosecond findings.\cite{Kastner.2019} It should be noted that the REMPI spectrum presented in Figure \ref{fig:REMPIvib} and Figure \ref{fig:PECDCurve} has been measured in a previous study at significantly reduced laser pulse power and was recorded on the parent ion mass.\cite{Kastner.2017} An increase in laser intensity leads to increased fragmentation but no change in the PECD as has been observed in a previous fs experiment.\cite{Lux.2015} In neither study was the PES or PECD strongly influenced by the fragmentation of the molecules. We thereby conclude that fragmentation of the molecules happens after the ionization in agreement with previous findings.\cite{Kastner.2019, Lux.2015} In a recent coincidence experiment on methyloxirane, a link between PECD and the fragmentation  channels in strong field ionization has been reported.\cite{Fehre.2019}

\subsection{PECD Evaluation}
\begin{figure}[tb]
\centering
\includegraphics[width=0.95\linewidth]{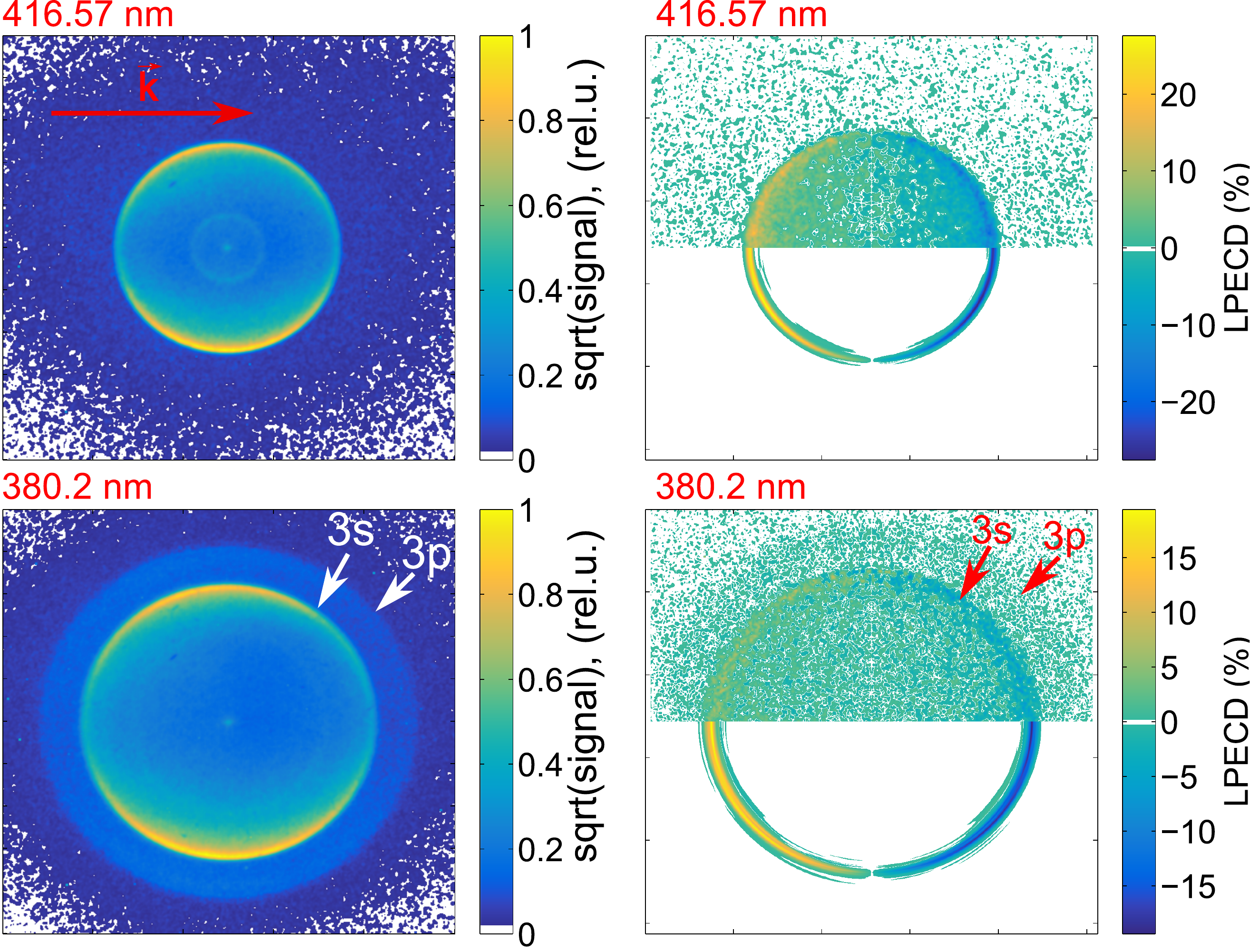}
\caption{\label{fig:PECD2}
Photoelectron velocity map images for (S)-($+$)-fenchone ionized via $2+1$ REMPI via the 0$_0^0$ transition to the 3s state using 416.57 nm light (top row) and via transitions to both the 3s and 3p states using 380.2 nm light (bottom row). The left-hand column shows the photoelectron images obtained using LIN and the right-hand column shows the antisymmetrized PECD images. In the upper half of each PECD image, the raw data is shown and in the lower half the Abel-inverted image is shown. In the images obtained with 380.2 nm light, the kinetic energy of the intense inner ring and weak outer ring correspond to vertical ionization from the 3s and 3p Rydberg states, respectively. 
}
\end{figure}

We calculate the raw PECD image at each wavelength by subtracting the RCP PAD image from the LCP PAD image.\cite{Lux.2015} In Figure \ref{fig:PECD2} we show the antisymmetric part of the raw and Abel-inverted PECD image for the $0_0^0$ transition at 416.57 nm and for excitation at 380.2~nm, where ionization via either the 3s or 3p states is possible. We observe a distinct forward/backward asymmetry of emitted photoelectrons with respect to the propagation axis of light.

\begin{figure}[tb]
\centering
\includegraphics[width=\linewidth]
{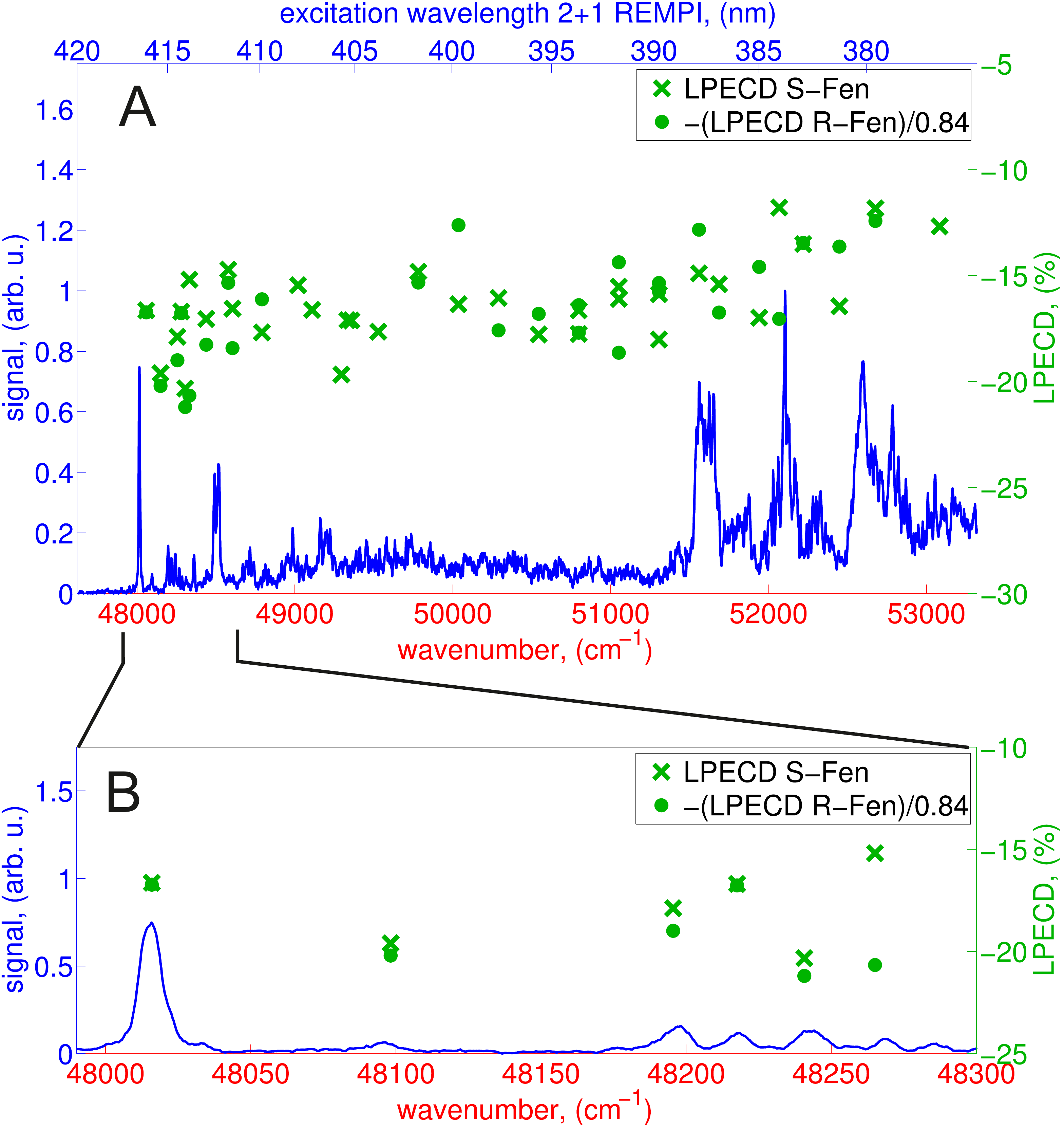}
\caption{\label{fig:PECDCurve}
A) High-resolution 2+1 REMPI spectrum showing partially resolved vibrational structure in the intermediate state (blue curve) and LPECD values for both enantiomers (green circles and crosses) for the 3s Rydberg state. Based on multiple measurements taken at a fixed frequency we deduce a relative experimental error of $\pm10$\%. The blue y-axis on the left side refers to the REMPI spectrum. The green y-axis on the right side refers to the LPECD values of (S)-($+$)-fenchone and (R)-($-$)-fenchone. As indicated in the legend, the LPECD values of (R)-($-$)-fenchone have been multiplied by $-\frac{1}{0.84}$. B) Enlarged view of the 3s band origin.
}
\end{figure}

From the raw PECD images, we calculate LPECD values for ionization via the 3s Rydberg state at each selected wavelength as described in Section \ref{sec:ExpSetup}. The results for both enantiomers are shown in Figure \ref{fig:PECDCurve} alongside the high-resolution REMPI spectrum of fenchone. Within the error of our experiment we observe the same magnitude (but opposite sign) for the two enantiomers taking into account the previously determined\cite{Kastner.2016} enantiomeric purity of 84\% of (R)-($-$)-fenchone. The LPECD decreases slightly in magnitude at shorter wavelengths. We could not extract reliable LPECD values for ionization from the 3p Rydberg state due to poor signal-to-noise ratio (see Figure \ref{fig:PECD2}).  

The LPECD values recorded after nanosecond laser ionization are larger in magnitude than the LPECD values recorded after femtosecond laser ionization\cite{Kastner.2017} over the measured energy range. The wavelength-dependent LPECD curves of the two measurements exhibit the same shape but are shifted vertically relative to one another by a value of ca. 0.025. However, it should be noted that the femtosecond laser LPECD values were recorded using an effusive beam source, whereas the nanosecond laser LPECD values were recorded using a pulsed molecular beam. Hence, vibrational and rotational excitation in the electronic ground state was greater in the femtosecond laser experiment and caution should be taken in making a direct comparison of the results. \\ \indent
As described in the previous section, we distinguish different regions of the spectrum in which the molecular dynamics are different. First, at wavelengths between 416.57\,nm and 412\,nm the density of vibrational levels in the 3s state is low so that we ionize fenchone essentially out of a true vibrational eigenstates. Second, at wavelengths between 412\,nm and 388\,nm, the density of states (DOS) is high, so that we ionize molecules whose vibrational energy is redistributed over a large number of coupled vibrational levels of the 3s state after IVR. Mode selective IC to the electronic ground state could also influence the vibrational state distribution.\cite{Comby.2016} Third, at wavelengths shorter than 388\,nm we excite both the 3s and the 3p states. However, internal conversion via 3p dominates the mechanism for ionization via 3s. Hence, the zero-order 3s vibrational state distribution is governed by vibronic coupling between initially populated vibrational levels of the 3p state and vibrational levels of the 3s state. \\ \indent
When comparing the LPECD values in the distinct spectral regions described above, we do not observe any abrupt changes of LPECD values. Furthermore, we do not observe any pronounced features in the LPECD values that correspond with resolved vibrational bands in the REMPI spectrum at the 3s origin between 416.57\,nm and 412\,nm. The LPECD value varies by ca.\ $\pm$3\% (absolute) when ionizing via different vibrational states, which is at the level of the experimental error of this measurement. \\ \indent
We therefore conclude that the vibrational energy of the intermediate 3s Rydberg state does not significantly influence the PECD of fenchone after 2+1 REMPI. This perhaps reflects the Ryd\-berg character of the intermediate electronic state. The electron's probability density is located relatively far from the molecular frame prior to ionization. The movement of the nuclei might therefore have only minor influence on the outgoing electron. This could be advantageous for future experiments as vibrational levels of the intermediate Rydberg state can be treated as spectator states, which do not influence the photoelectron angular distribution or energy in experiments on a nanosecond timescale. However, further experiments on different chiral molecules need to be done to clarify whether these observations can be generalized.

\section{Summary and Conclusion}
\label{sec:Summ}
We used nanosecond laser pulses to obtain the photoelectron circular dichroism (PECD) of the chiral molecule fenchone after 2+1 resonance enhanced multi-photon ionization (REMPI) via the 3s and 3p Rydberg states for selected ionization energies in the spectral range between 375 nm and 420 nm. In contrast to previous studies using femtosecond laser ionization, we resolve vibrational structure in the REMPI spectrum, and assign LPECD values to distinct intermediate vibrational levels of the 3s Rydberg state. We do not observe any strong dependence of the LPECD value on the vibrational level of the intermediate state, which might be a consequence of its Rydberg electronic character. An accompanying computational study identifies the different vibrational modes that give rise to the peaks observed in the 2+1 REMPI spectrum. We demonstrate the feasibility of PECD measurements with tabletop nanosecond dye laser systems, which can be used to investigate chiral mixtures via high resolution REMPI assisted PECD measurements. 

\section*{Acknowledgement}

The authors are grateful to Theofanis Kitsopoulos for support and feedback. Funded by the Deutsche Forschungsgemeinschaft (DFG, German Research Foundation) – Projektnummer 328961117 – SFB 1319 and ELCH. A.K, G.B.P. and T.S. acknowledge financial support within the LaserLab Europe network (Grant Agreement No. 654148). T.S. acknowledges support by the DFG under grant INST 186/1302-1. P.S. gratefully acknowledges support by HELLAS-CH (MIS 5002735) implemented under “Action for Strengthening Research and Innovation Infrastructures”, funded by the Operational Programme “Competitiveness, Entrepreneurship and Innovation” (NSRF 2014-2020) and co-financed by Greece and the European Union (European Regional Development Fund).

\bibliography{VibPECD_bib}
\bibliographystyle{rsc} 

\providecommand*{\mcitethebibliography}{\thebibliography}
\csname @ifundefined\endcsname{endmcitethebibliography}
{\let\endmcitethebibliography\endthebibliography}{}

\end{document}